%% file: ms.tex
\documentclass[apj]{emulateapj}
\usepackage{amsmath}
\usepackage{apjfonts}
\usepackage{graphicx}
\input{Definitions.tex}
\begin{document}
\newcommand{\vv}{\textrm{v}}
\title{The energy production rate \& the generation spectrum of UHECRs}
\author{Boaz Katz\altaffilmark{1}, Ran Budnik\altaffilmark{1} and Eli
Waxman\altaffilmark{1}}
\altaffiltext{1}{Physics Faculty, Weizmann Institute, Rehovot 76100, Israel; boazka@wizemail.weizmann.ac.il, waxman@wicc.weizmann.ac.il}
\begin{abstract}
We derive simple analytic expressions for the flux and spectrum of ultra-high energy cosmic-rays (UHECRs) predicted in models where the CRs are protons produced by extra-Galactic sources. For a power-law scaling of the CR production rate with redshift and energy, $d\dot{n}/d\vep\propto\vep^{-\alpha}(1+z)^m$, our results are accurate at high energy, $\vep>10^{18.7}$~eV, to better than $15\%$, providing a simple and straightforward method for inferring $d\dot{n}/d\vep(\vep)$ from the observed flux at $\vep$. We show that current measurements of the UHECR spectrum, including the latest Auger data, imply  $\vep^2d\dot{n}/d\vep(z=0)=(0.45\pm0.15)(\alpha-1)\times10^{44}\erg\Mpc^{-3}\yr^{-1}$ at $\vep>10^{19.5}\text{eV}$ with $\alpha$ roughly confined to $2\lesssim\alpha<2.7$. The uncertainty is dominated by the systematic and statistic errors in the experimental determination of individual CR event energy, $(\Delta\vep/\vep)_{\rm sys}\sim(\Delta\vep/\vep)_{\rm stat}\sim20\%$. At lower energy, $d\dot{n}/d\vep$ is uncertain due to the unknown Galactic contribution. Simple models in which $\alpha\simeq2$ and the transition from Galactic to extra-Galactic sources takes place at the "ankle", $\vep\sim 10^{19}\eV$, are consistent with the data. Models in which the transition occurs at lower energies require a high degree of fine tuning and a steep spectrum, $\alpha\simeq2.7$, which is disfavored by the data. We point out that in the absence of accurate composition measurements, the (all particle) energy spectrum alone cannot be used to infer the detailed spectral shapes of the Galactic and extra-Galactic contributions.
\end{abstract}

\section{Introduction}
The origin of the observed Cosmic Rays (CRs) at different energies is still unknown \citep[see][for
reviews]{Blandford87,Axford94,Nagano00}. The cosmic ray spectrum changes its qualitative behavior as a function of
particle energy; it steepens around $\sim 5\times 10^{15}$~eV (the ``knee'') and flattens around $5\times 10^{18}$~eV
(the ``ankle'').  Below $\sim 10^{15}$~eV, the cosmic rays are thought to originate from Galactic supernovae. The
composition is dominated by protons at the lowest energies, and the fraction of heavy nuclei increases with energy. The
proton fraction at $\sim 10^{15}$~eV is reduced to $\sim15\%$ \citep{Burnett90,Bernlohr98}. At yet higher energies, there
is evidence that the fraction of light nuclei increases, and that the cosmic-ray flux above $5\times 10^{18}$~eV is again
dominated by protons (Gaisser et. al 1998; Bird et al. 1994; note that preliminary results by the Auger collaboration
suggest that at the highest energies the composition becomes heavier again, e.g. Bluemer; for the Pierre Auger
Collaboration 2008). The composition change and the flattening of the spectrum around $10^{19}$~eV (see Fig.
\ref{fig:ReasonableFit}) suggest that the flux above and below this energy is dominated by different sources. At energies
of $\vep_{19}\equiv\vep/10^{19}\eV\sim1$ the Larmor radius of CRs in the Galactic magnetic field is
\begin{equation}
R_{L}\sim 3 B_{-5.5}^{-1}\vep_{19}Z^{-1}\kpc,
\end{equation}
where $B=3B_{-5.5}\mu G$ is the value of the Galactic magnetic field and $Z$ is the charge of the observed Galactic cosmic rays. Since
the Galactic magnetic field can not confine protons above $10^{19}$~eV, it is believed that the nearly isotropic cosmic ray flux
at $\varepsilon>5\times 10^{18}$~eV originates from extra-Galactic (XG) sources. The small, but statistically
significant, enhancement of the flux at $\varepsilon<3\times 10^{18}$~eV near the Galactic plane \citep{Bird99,Hayashida99},
suggests a Galactic origin at these lower energies. Note, however, that the Auger experiment did not detect any anisotropy in
the energy regime $10^{18}\eV\lesssim\vep<3\times 10^{18}\eV$ \citep[][and references therein]{Leuthold08}, in conflict
with earlier results.

CRs with energies exceeding $\varepsilon>5\times 10^{18}$~eV are termed Ultra High Energy Cosmic Rays (UHECRs). The
(probably XG) sources of these particles are unknown and have been the issue of much debate \citep[e.g.][and references
therein]{Waxman04,Berezinsky08}. Measurements of the fluxes of particles in this energy range have been conducted by
several groups including AGASA \citep{Takeda98}, Fly's Eye \citep{Bird94}, HiRes \citep{Abbasi08}, Yakutsk
\citep{Afanasiev93} and most recently Auger \citep{Bluemer08}.

The problem of inferring the generation rate and generation spectrum of UHECRs from the observed flux and spectrum has been addressed by several authors \citep[e.g. ][and references therein]{Waxman95b, Bahcall03,Berezinsky06,Berezinsky08,Stanev08}. As long as we restrict to the highest UHECRs the different models roughly agree with each other. For example, \citet{Waxman95b} finds a generation spectrum $d\dot n/d\vep\equiv Q(\vep)\propto \vep^{-\alpha}$ with $\alpha=-2.3\pm0.5$ and energy production rate of CRs in the energy range $10^{19}\eV<\vep<10^{21}\eV$ of $4.5\pm1.5\times10^{44}\erg\Mpc^{-3}\yr^{-1}$, while in the model presented by \citet{Berezinsky06}, the spectral index is $\alpha=2.7$ and the energy generation rate in the energies $10^{19}\eV<\vep<10^{21}\eV$ is approximately $6\times 10^{44}\erg\Mpc^{-3}\yr^{-1}$.  The main difference between different models is the location of the transition between Galactic dominated and XG dominated CRs.
While some models assume that the transition is at energies of $\vep\sim 10^{19}\eV$ \citep[e.g. ][]{Waxman95b}, other models alow for a transition at considerably lower energies, $\vep\lesssim 10^{18}\eV$ \citep[e.g. ][]{Berezinsky06}.

In this paper we revisit this problem, presenting a novel analytic tool for the analysis of the UHECR spectrum and
including the most recent data from the Auger experiment. Our estimates for the energy generation rate, are in agreement with previous estimates (see \sref{sec:Production}).

This paper is organized as follows. We first describe the framework of our analysis in \sref{sec:Model}. We focus
on models where the UHECRs are dominated by protons originating from XG sources with power-law generation spectrum and an
energy production rate that depends on redshift. We derive simple analytic approximations for the calculation of the
effects of the energy losses due to interaction with the CMB, and show that the approximate analytic results reproduce
with good accuracy the results of direct numerical calculations for particle energies of $\vep\gtrsim 10^{18.7}\eV$.
Next, we estimate in \sref{sec:Production} the energy production rate of cosmic rays with energies
$\varepsilon>10^{19}\eV$, and show that it is roughly model independent. In section \sref{sec:Transition} we show that
the measured spectrum of CRs at energies $\vep>10^{16}\eV$ is consistent with simple models with a transition between Galactic and XG dominated regions at the 'ankle'. We review and illustrate the argument that models in which the transition
between Galactic and XG sources occurs at energies considerably below the ankle require a high degree of fine tuning.
We summarize the results and the main conclusions in \sref{sec:Discussion}.

\section{Analytic CR propagation model}\label{sec:Model}
In this section we derive approximate analytic  expressions for the expected UHECR energy dependent flux assuming it is
dominated by XG proton sources. First we derive in \sref{sec:AnalyticProp} approximate expressions for the flux in terms
of the energy loss time $\tau$ by assuming that at high energies $\tau$ is much shorter than the Hubble time $H_0^{-1}$.
The approximate expressions are derived by calculating the effect of the losses to first order in the small parameter
$H_0\tau$. We then obtain in \sref{sec:AnalyticCMB} a phenomenological analytic approximation for the loss time of
protons due to pair production and pion production and use it to derive an analytic expression for the expected flux. We
show that these expressions are in good agreement with detailed numerical calculations.

\subsection{Analytic estimate of the effects of CR energy losses on the observed spectrum}\label{sec:AnalyticProp}
As they propagate, high-energy protons lose energy as a result of the cosmological redshift and as a result of production
of pions and $e+e-$ pairs in interactions with CMB photons. Here we approximate the energy loss caused by scattering,
which is a random process, as a continuous energy loss (CEL). In this approximation, the energy loss rate of a proton of
given energy is taken to be the mean loss rate of an ensemble of protons of the same energy. The CEL approximation is
excellent for pair production, for which the mean free path is small and the relative energy loss in a single scattering,
of order $m_e/m_p\sim10^{-3}$, is also small. For pion production, the average relative energy loss in a single collision
is 0.13 at the threshold and rises to 0.5 at higher energy. However, fluctuations in proton energy resulting from this
process are significant only for propagation distances smaller than 100 Mpc and proton energies $>10^{20}$ eV
\citep{Aharonian94}. The CEL approximation therefore gives accurate results for the flux below $10^{20}$ eV. At higher
energies, the flux obtained by this approximation drops faster than that obtained when fluctuations in proton energy are
taken into account. It has been shown \citep{Berezinsky75,Berezinskii88} that for a flat generation spectrum,
$\alpha<2.6$, the flux obtained using the CEL approximation is accurate to better than $10\%$ up to $3\times10^{20}$ eV,
the highest energy at which events have been reported.

For numerical estimates we assume the following cosmological parameters: $H_0=72\km\sec^{-1}\Mpc^{-1}$, $\Omega_{M}=0.28$
and $\Omega_L=0.72$. We assume that the generation spectrum is a power law and that the production rate depends on $z$ as
\begin{equation}\label{eq:Q}
d\dot{n}/d\vep\equiv Q(\vep,z)=(\vep/\vep_0)^{-\alpha}(1+z)^mQ(\vep_0)
\end{equation}
for $z<z_{max}$ and $\vep<\vep_{max}$, where $z_{max}$ and $\vep_{max}$ represent the maximal $z$ and particle energy
respectively, up to which CRs are generated. In all numerical calculations we adopt $z_{\max}=4$ and
$\vep_{\max}=10^{22}\eV$. For energies $\vep>10^{18.7}\eV$ the results are insensitive to these choices. In all numerical
calculations we use the loss time calculated by \citet{Berezinsky06}.

Consider a CR that propagated through the IGM while suffering energy losses due to interactions with the CMB and due to
the expansion of the universe and reached us at $t=0$ with an energy $\vep_0$. Denote the energy that it had at an
earlier time $t$ by $\vep'(\vep_0,t)$. The time dependent energy $\vep'$ satisfies the equation:
\begin{equation}
\pr_t \vep'(\vep_0,t)=\left(\tau_{CMB}^{-1}(\vep',t)+\frac{1}{1+z}\frac{dz}{dt}\right)\vep',
\end{equation}
where $\tau_{CMB}(\vep',t)$ is the energy loss time due to interactions with the CMB and the second term on the rhs is
the energy loss rate due to the Hubble expansion. The spectrum of particles observed at the current time is given by:
\begin{equation}
\frac{dn}{d\vep}(\vep_0)d\vep_0=\int_{-\infty}^{0} dt Q(\vep'(\vep_0,t),t)\frac{\pr\vep'}{\pr\vep_0}|_td\vep_0
\end{equation}
\begin{equation}
=\int_{\vep_0}^{\infty} \frac{d\vep'}{\vep'\tau^{-1}(\vep',t)} Q(\vep'(\vep_0,t),t)\frac{\pr\vep'}{\pr\vep_0}|_t d\vep_0,
\end{equation}
where $Q$ is the CR generation rate per comoving volume per unit energy.

At high energies the loss time due to the interactions with the CMB is much shorter than the Hubble time: $\tau_{CMB}\ll
H_0^{-1}$. In what follows, we use the assumption that $H_0\tau_{CMB}\ll1$ to find an approximate relation between the
observed spectrum and the CR production spectrum.

The CMB spectrum scales with redshift as $n_{CMB}(\vep,z)=(1+z)^3 n_{CMB}(\vep/(1+z),z)$ and thus all continuous energy
losses due to interactions with the CMB have the following scaling,
\begin{equation}\label{eq:tau_scaling}
\tau_{CMB}(\vep,z)=(1+z)^{-3}\tau_{CMB}((1+z)\vep).
\end{equation}
It is useful to consider the evolution of the scaled energy $\tilde\vep=(1+z)\vep$. We find
\begin{equation}
\inv{\tilde\vep}\frac{d\tilde\vep}{dt}=\tau_{CMB}^{-1}(\vep,z)+\frac{2}{1+z}\frac{dz}{dt}\approx(1+z)^3\tau_{CMB}^{-1}(\tilde\vep)(1+2H_0\tau_{CMB}(\tilde\vep)),
\end{equation}
where we neglected second order terms in $H_0\tau_{CMB}(\tilde\vep)$.

Next, we replace the time variable with a time parameter $\tilde t$ that satisfies $d\tilde t=(1+z)^3dt$ and obtain
\begin{equation}
\inv{\tilde\vep}\frac{d\tilde\vep}{d\tilde t}\approx\tau_{CMB}^{-1}(\tilde\vep)(1+2H_0\tau_{CMB}(\tilde\vep)).
\end{equation}
Since the time derivative of $\tilde\vep$ does not depend on time, we can write
\begin{equation}
\frac{dn}{d\vep}(\vep_0)d\vep_0=d\tilde t_0 \int_{\tilde\vep_0}^{\infty} d\tilde\vep \frac{\pr \vep \pr t}{\pr \tilde\vep
\pr \tilde t}Q(\vep,t),
\end{equation}
where
\begin{equation}
d\tilde t_0=d\vep_0\frac{\tau_{CMB}(\vep_0)}{1+2H_0\tau_{CMB}(\vep_0)}.
\end{equation}
Since $\tilde t$ is a function of $t$ only, we have
\begin{equation}
\frac{\pr \vep \pr t}{\pr \tilde\vep \pr \tilde t}=\frac{\pr \vep}{\pr \tilde\vep}\frac{\pr t}{\pr \tilde t}=(1+z)^{-4},
\end{equation}
and thus
\begin{equation}\label{eq:general_first_order}
\frac{dn}{d\vep}(\vep_0)\approx\frac{\tau_{CMB}(\vep_0)}{1+2H_0\tau_{CMB}(\vep_0)}\int_{\vep_0}^{\infty} (1+z)^{-4}Q(\frac{\tilde\vep}{1+z},z)d\tilde\vep_0.
\end{equation}

For a generation spectrum $Q(\vep,z)=(\vep/\vep_0)^{-\alpha}(1+z)^mQ(\vep_0)$ we find
\begin{equation}\label{eq:lastdnde}
\frac{dn}{d\vep}(\vep_0)\approx\frac{\tau_{CMB}(\vep_0)}{1+2H_0\tau_{CMB}(\vep_0)}Q(\vep_0)\int_{\vep_0}^{\tilde\vep_{\max}}
(1+z)^{-4+m+\alpha}\left(\frac{\tilde\vep}{\vep_0}\right)^{-\alpha}.
\end{equation}

It is useful to express the relation between the generation spectrum and the observed flux in terms of an effective CR
generation time $t_{\text{eff}}$ defined by
\begin{equation}
\frac{dn}{d\vep}(\vep_0)\equiv Q(\vep_0)t_{\text{eff}}(\vep_0).
\end{equation}
Equation \eqref{eq:lastdnde} can then be written as
\begin{equation}
t_{\text{eff}}(\vep_0)\approx\frac{\tau_{CMB}(\vep_0)}{1+2H_0\tau_{CMB}(\vep_0)}\int_{\vep_0}^{\vep_{\max}}
(1+z)^{-4+m+\alpha}\left(\frac{\vep}{\vep_0}\right)^{-\alpha},
\end{equation}
where we assumed that replacing $\tilde\vep_{\max}$ with $\vep_{max}$ in the integration upper limit results in a small
error which can be considered second order.
For the special case $\alpha+m=4$ we find that
\begin{equation}\label{eq:t_th_m_alpha_4}
t_{\text{eff}}(\vep_0)\approx\left(\inv{\alpha-1}\right)\frac{\tau_{CMB}(\vep_0)}{1+2H_0\tau_{CMB}(\vep_0)}\left[1-\left(\frac{\vep_{\max}}{\vep_0}\right)^{1-\alpha}\right].
\end{equation}
For most practical purposes, assuming that $\vep_{max}$ is large enough, we can ignore the last term on the rhs of Eq.
\eqref{eq:t_th_m_alpha_4} to obtain
\begin{equation}\label{eq:t_th_m_alpha_4_short}
t_{\text{eff}}(\vep_0)\approx\left(\inv{\alpha-1}\right)\frac{\tau_{CMB}(\vep_0)}{1+2H_0\tau_{CMB}(\vep_0)}.
\end{equation}

\begin{figure}[h]
\epsscale{1.3} \plotone{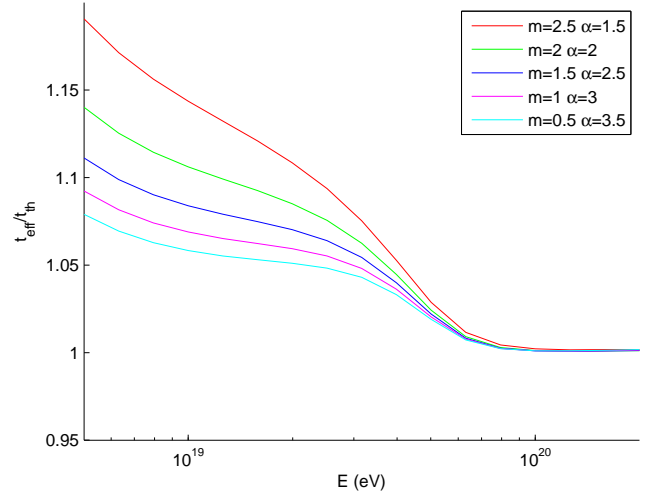} \caption{Comparison of equation \eqref{eq:t_th_m_alpha_4} with direct numerical
integration of the propagation equations for $\alpha+m=4$. We use the numerical estimates for the energy loss time,
$\tau_{CMB}$, given by \citet{Berezinsky06}.}\label{fig:t_th_m_alpha_4}
\end{figure}

We see that in case $\alpha+m=4$ the effective generation time depends to first order only on the value of $\tau_{CMB}$
at $\vep_0$. In figure \ref{fig:t_th_m_alpha_4} we compare the results of numerical calculations of the propagation of
CRs to what we expect based on equation \eqref{eq:t_th_m_alpha_4} for different values of $\alpha$ and $m$ that satisfy
$\alpha+m=4$. As can be seen in the figure, for $\alpha>2$, the differences at large energies $\vep\gg5\times10^{18}$ are smaller than
$\sim 15\%$.

In case $\alpha+m\neq4$ we need to evaluate $z(\tilde\vep,\vep_0)$. To first order in $H_0\tau_{CMB}$ we have:
\begin{equation}
dz\approx H_0d\tilde t\approx H_0\tau_{CMB}(\tilde\vep)\frac{d\tilde\vep}{\tilde\vep}
\end{equation}
so
\begin{equation}\label{eq:z_tilde_vep}
z\approx \int_{\vep_0}^{\tilde\vep} H_0\tau_{CMB}(\vep)\frac{d\vep}{\vep}.
\end{equation}

\begin{figure}[h]
\epsscale{1.3} \plotone{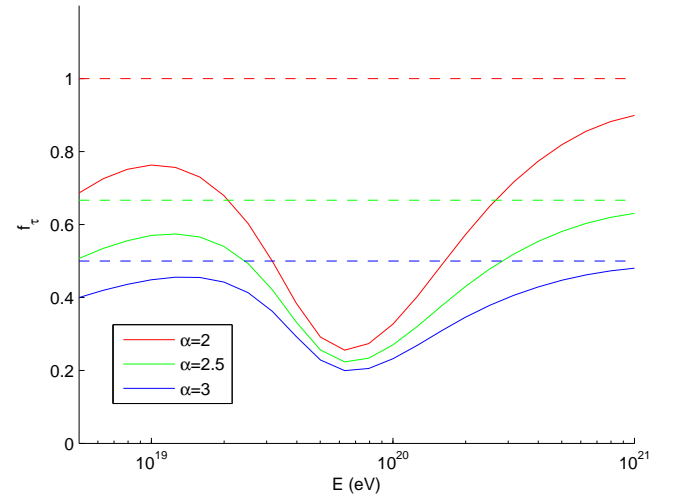} \caption{Numerical values of $f_{\tau}(\alpha,\vep)$, given by Eq.
\eqref{eq:f_tau}, for energy losses of protons interacting with the CMB (full lines). Dashed lines are the upper limits
$1/(\alpha-1)$.}\label{fig:f_tau}
\end{figure}

\begin{figure}[h]
\epsscale{1.3} \plotone{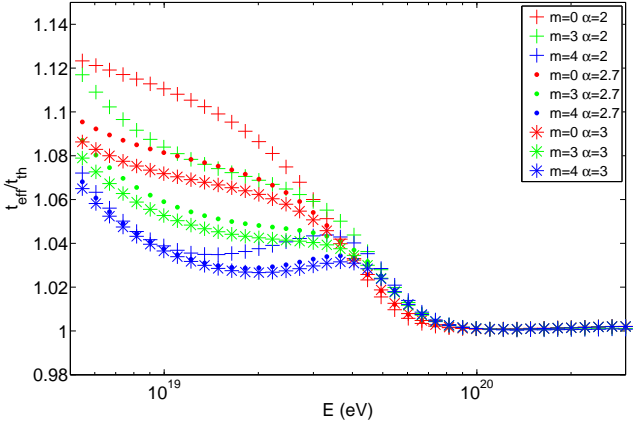} \caption{Comparison of equation \eqref{eq:t_th} with direct numerical
calculations for $\alpha+m\neq4$.}\label{fig:t_th}
\end{figure}

Using equations \eqref{eq:z_tilde_vep} and \eqref{eq:general_first_order}, we find that to first order in $H_0\tau_0$
(after some straight-forward algebraic manipulation)
\begin{equation}\label{eq:t_th}
t_{\text{eff}}(\vep_0)\approx\frac{\tau_{CMB}}{\alpha-1}\left[1+H_0\tau_{CMB}(\vep_0)\left(2+(4-\alpha-m)f_{\tau}(\vep_0,\alpha)\right)\right]^{-1},
\end{equation}
where
\begin{equation}\label{eq:f_tau}
f_{\tau}(\alpha)=\int_1^{\infty}dx\frac{\tau_{CMB,0}(x\vep_0)}{\tau_{CMB}(\vep_0)}x^{-\alpha}.
\end{equation}
Assuming that $\tau_{CMB}(\vep)$ is a decreasing function of energy, we have
\begin{equation}
0<f(\alpha)<\frac{1}{\alpha-1}.
\end{equation}
The values of $f_{\tau}$ for different values of $\alpha$ and different energies are presented in figure \ref{fig:f_tau}
(full lines). In this figure, the values of the upper limit $1/(\alpha-1)$ are shown as dashed lines. In figure
\ref{fig:t_th} we compare the results of numerical calculations of the propagation of CRs to the results of equation
\eqref{eq:t_th}. As can be seen in the figure, at large energies, $\vep>5\times10^{18}$, Eq. \eqref{eq:t_th} is accurate
to better than $15\%$.

\subsection{Energy losses of proton CRs due to interactions with the CMB}\label{sec:AnalyticCMB}
The loss time of protons due to interactions with the CMB that produce pairs and pions has been calculated in several
publications. These processes are characterized by the existence of a threshold energy of the photons in the rest frame
of the protons, that is required in order to create the products of the interaction. The typical energies required from a
proton in order to produce an electron positron pair and a pion by interacting with a photon with typical energy
$3T_{CMB}$ are roughly
\begin{equation}\label{eq:ep_threshold_3T}
\vep_{th,ep}\sim 2m_{p}m_{e}c^4/(3T_{CMB})\sim 1.4 \times 10^{18}\eV
\end{equation}
and
\begin{equation}\label{eq:pion_threshold_3T}
\vep_{th,\pi}=m_{p}m_{\pi}c^4/(3T_{CMB})\sim 2\times 10^{20}\eV
\end{equation}
respectively.

For protons with energies that are much smaller, photons with energies that are much larger are required in order to
create the products. The numbers of contributing photons decreases exponentially as the proton energy is decreased.
Motivated by this fact, we use the following anzatz \citep{Waxman95b} to parameterize the energy loss time:
\begin{align}\label{eq:proton_loss_approx}
\tau^{-1}(\vep)=\tau_{0,ep}^{-1} \exp(-\vep_{c,ep}/\vep)+\tau_{0,\pi}^{-1} \exp(-\vep_{c,\pi}/\vep),
\end{align}
where the values of $\tau_{0,ep/\pi}$ and $\vep_{c,ep/\pi}$ are chosen to fit numerical calculations of the loss time. By
setting $\vep_{c,ep}=2.7\times10^{18}\eV$, $\tau_{0,ep}=3.4\times10^9\yr$;  $\vep_{c,\pi}=3.2\times10^{20}\eV$ and
$\tau_{0,\pi}=2.2\times10^7\yr$, equation \eqref{eq:proton_loss_approx} reproduces the results of \citet{Berezinsky06} to
better than $10\%$ accuracy in the energy range $10^{18.7}<\vep<10^{20.5}$ (this approximation for the pion production
agrees with that of \citet{Waxman95b} to better than 10\%).

The values for  $\vep_{c,ep}$ and $\vep_{c,\pi}$ are of the same order of magnitude as the threshold energies in
\eqref{eq:pion_threshold_3T} and \eqref{eq:ep_threshold_3T}. To get a feeling for where the typical values for
$\tau_{ep}$ and $\tau_{\pi}$ come from, we can make crude estimates as follows: For both processes we express the typical
loss time in terms of a typical cross section and relative energy loss
\begin{equation}
\tau\sim (n_{CMB}\sig c\eta^{-1})^{-1},
\end{equation}
where for electron positron production we have roughly $\sig_{ep}\sim\alpha_e\sig_T$ and $\eta_{ep}\sim m_e/m_p$ so
\begin{equation}\label{eq:ep_tau0}
\tau_{ep}\sim (n_{CMB}\alpha_e\sig_T cm_p/m_e)^{-1}\sim 10^9\yr,
\end{equation}
while for pion production we have roughly $\sig_{\pi}\sim (\hbar/m_\Delta c)^2$ and $\eta_{\pi}\sim m_{\pi}/m_p$
so
\begin{equation}\label{eq:pion_tau0}
\tau_{\pi}\sim (n_{CMB}(\hbar/m_\Delta c)^2m_p/m_{\pi})^{-1}\sim 7\times 10^7\yr.
\end{equation}
We see that the crude estimates in equations \eqref{eq:pion_tau0} and \eqref{eq:ep_tau0}, are of the same order of magnitude as the fitted values.

We can get a better fit by using more parameters. Using the following anzatz:
\begin{align}\label{eq:proton_loss_approx_better}
\tau^{-1}(\vep)=(\tau_{0,ep}\exp(\vep_{c,ep}/\vep)+d\tau_{ep})^{-1}+(\tau_{0,\pi}\exp(\vep_{c,\pi}/\vep)+d\tau_{\pi})^{-1}
\end{align}
and setting the values $\vep_{c,ep}=9.1\times10^{18}\eV$, $\tau_{0,ep}=0.5\times10^9\yr$, $d\tau_{ep}=3\times10^9\yr$;
$\vep_{c,\pi}=3.5\times10^{20}\eV$,  $\tau_{0,\pi}=1.4\times10^7\yr$, and $d\tau_{\pi}=2.4\times10^9\yr$, equation
\eqref{eq:proton_loss_approx_better} reproduces the results of \citet{Berezinsky06} to better than $3\%$ accuracy in the
energy range $10^{18.7}<\vep<10^{20.5}$.

For most purposes it is sufficient to consider only the simple cases where $\alpha+m=4$, and use the approximate
expression given in \eqref{eq:t_th_m_alpha_4_short}. Using equation \eqref{eq:proton_loss_approx} we get
\begin{equation}\label{eq:proton_effective_time}
(\alpha-1)t_{eff}(\vep)\approx \left[\tau_{0,ep}^{-1} \exp(-\vep_{c,ep}/\vep)+\tau_{0,\pi}^{-1}
\exp(-\vep_{c,\pi}/\vep)+2H_0\right]^{-1}.
\end{equation}

\begin{figure}[h]
\epsscale{1.3} \plotone{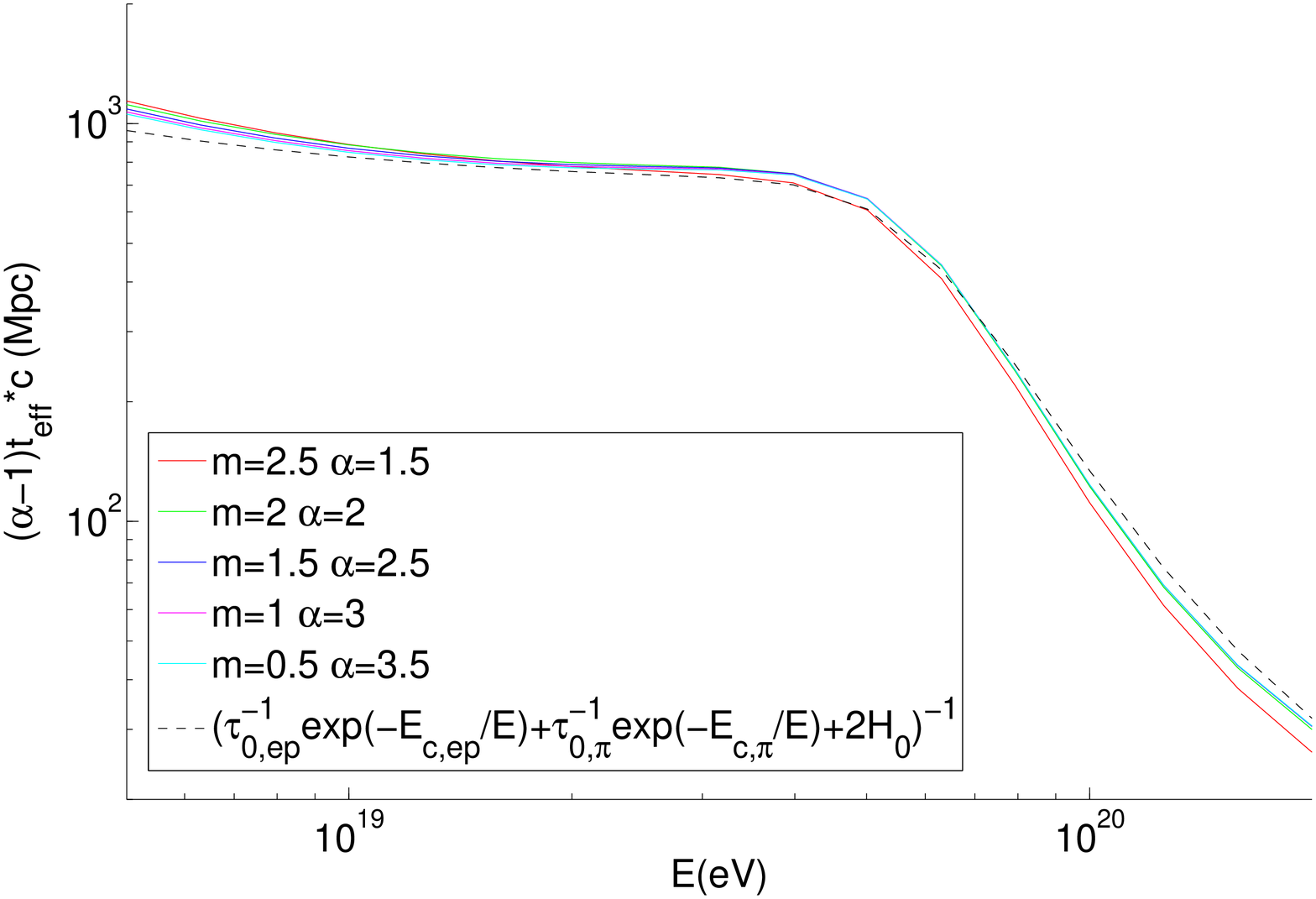} \caption{Comparison of direct numerical calculations of the effective
CR production time, with the analytic approximation \eqref{eq:proton_effective_time} for $\alpha+m=4$. The effective time
is multiplied by $\alpha-1$ to show the similarity of the energy dependence and to compare with the analytic expression.
The time is shown in units of Mpc.}\label{fig:t_th_m_alpha_4_analytic_Mpc}
\end{figure}

A comparison of numerical calculations of the effective time $t_{eff}$ with equation \eqref{eq:proton_effective_time} for
values of $\alpha$ and $m$ that satisfy $\alpha+m=4$ is shown in figure \ref{fig:t_th_m_alpha_4_analytic_Mpc}. The
effective time (multiplied by the speed of light) is shown in units of Mpc to show the effective distance from which these
CRs can reach us.

\begin{figure}[h]
\epsscale{1.3} \plotone{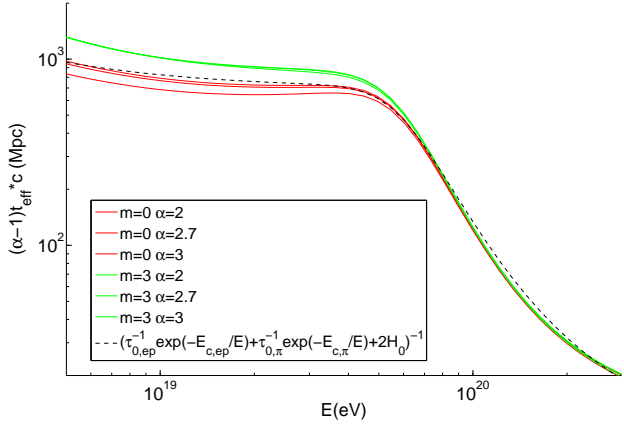} \caption{Comparison of direct numerical calculations of the effective CR
production time, with the analytic approximation \eqref{eq:proton_effective_time} for limited values of $m$ and $\alpha$.
The effective time is multiplied by $\alpha-1$ to show the similarity of the energy dependence and to compare with the
analytic expression. The time is shown in units of Mpc.}\label{fig:t_th_m_alpha_0_3_2_3_analytic_Mpc}
\end{figure}

In fact, for energies $\vep>10^{19}\eV$, using equation \eqref{eq:proton_effective_time}, gives a good approximation (to
better than $20\%$) to the effective time for values of $m$ and $\alpha$ in the range $2<\alpha<3$ and $0<m<3$ as shown
in figure \ref{fig:t_th_m_alpha_0_3_2_3_analytic_Mpc}.

\section{Energy production rate of $10^{19}\eV<\vep<10^{21}\eV$ UHECRs}\label{sec:Production}
\begin{figure}[h]
\epsscale{1.3} \plotone{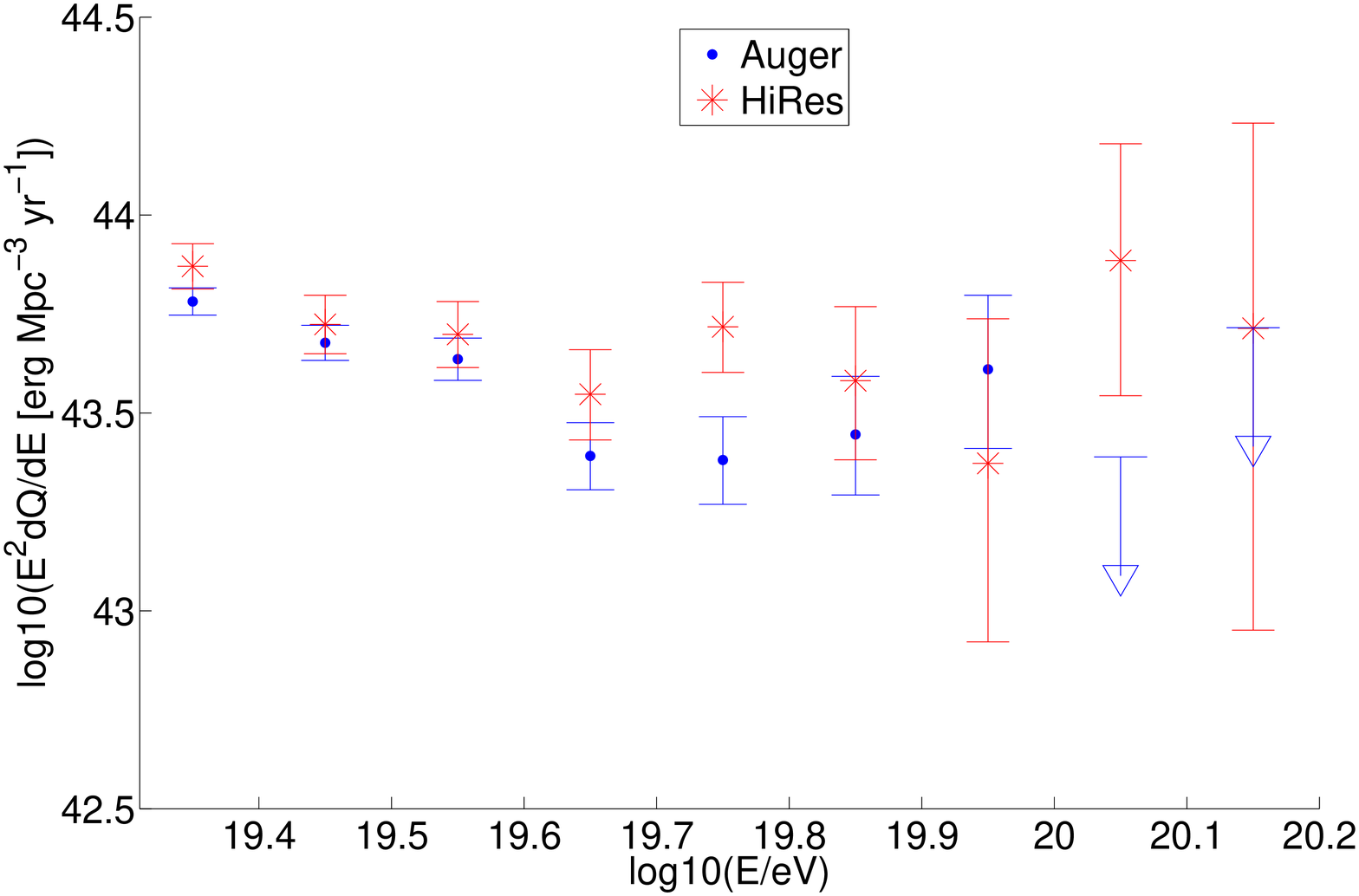} \caption{The energy generation rate as measured by Auger and Hires assuming
that the CRs are purely protons. We used equation \eqref{eq:proton_effective_time} setting $\alpha-1=1$. For $0<m<3$, Eq.
\eqref{eq:proton_effective_time} is accurate to better than $20\%$. Statistical and systematic errors in the experimental
determination of event energies lead to $\sim 50\%$ errors in the flux at the highest energies. For different values of
$\alpha$, the spectrum should be multiplied by an energy independent factor $(\alpha-1)$. The absolute energy scales of
the Auger and Hires data where not altered in this figure.}\label{fig:EnergyGeneration}
\end{figure}

In this section we discuss the energy production rate of UHECRs required to explain the observed CR flux. Assuming that
the CRs are mostly protons, we can use equation \eqref{eq:proton_effective_time} to estimate the effective generation
time of CRs in the energy range $\vep>10^{19}\eV$. The energy production of CRs per logarithmic energy interval can be
approximated by:
\begin{equation*}
\vep^2Q(\vep)\approx
\end{equation*}
\begin{equation}
(\alpha-1)\left[\tau_{0,ep}^{-1} \exp(-\vep_{c,ep}/\vep)+\tau_{0,\pi}^{-1}
\exp(-\vep_{c,\pi}/\vep)+2H_0\right]\vep^2\frac{dn}{d\vep},
\end{equation}
where $dn/d\vep=J/(4\pi c)$ is the measured CR particle density per unit energy. Assuming that the cosmic ray spectrum
index is in the range $2<\alpha<3$ and that $0<m<3$ [see Eq. \eqref{eq:Q}], this equation is accurate to better than 20\%
(as explained below, systematic and statistic experimental errors lead to a larger uncertainty in Q). The energy
generation implied by the latest data of Auger and Hires for $\vep>10^{19.3}\eV$ is shown in figure
\ref{fig:EnergyGeneration}. As can be seen in figure \ref{fig:EnergyGeneration}, the energy generation rate in the range
$\vep>10^{19.3}\eV$ is $\vep^2Q(\vep)\approx 10^{43.5}-10^{44}\erg\Mpc^{-3}\yr^{-1}$. It should be noted that some of the
CRs in this energy range might be of Galactic origin (especially in the lower energy part of this range) and some of them may
not be protons. In this respect, the energy production shown in the figure can be considered an upper limit to the
energy production in XG UHECR protons. In addition, at energies $\vep\gtrsim 10^{20}\eV$, the local CR source density,
which may be different from the average one, may affect the measured flux
\citep[e.g.][]{Giler80,Bahcall00,Waxman95b,Berezinsky08}.

The statistic and systematic errors in the generation rate estimate are of the order of tens of percents. The main
source for the systematic error is a $\sim 20\%$ systematic uncertainty in the energy calibration of the CR particle
energies, which implies a $\sim 30\%$ uncertainty in the measured flux. In addition, all experiments have statistic
errors in the particles' energy estimates. As shown in appendix \sref{sec:Smear}, the flux is overestimated by a factor
of [Eq. \eqref{eq:statistics_approx}]
\begin{equation}\label{eq:smear}
f_{\text{stat}}\approx \left[1+0.5(\beta-1)(\beta-2)\sig_{\text{stat}}^2\right],
\end{equation}
were $\sig_{\text{stat}}^2=\ave{(\Delta\vep_{\text{stat}}/\vep)^2}$ and $\beta=d\log J/d\log\vep$. For a value of
$\sig_{\text{stat}}\approx 0.2$ typical to these experiments \citep[e.g.][]{Abraham08} this implies an error of $30-40\%$
at peak ($\vep\sim 10^{20}\eV$). If higher precision is sought, an energy dependent estimate of $\sigma_{\text{stat}}$ at these
energies is essential.

\begin{figure}[h]
\epsscale{1.3} \plotone{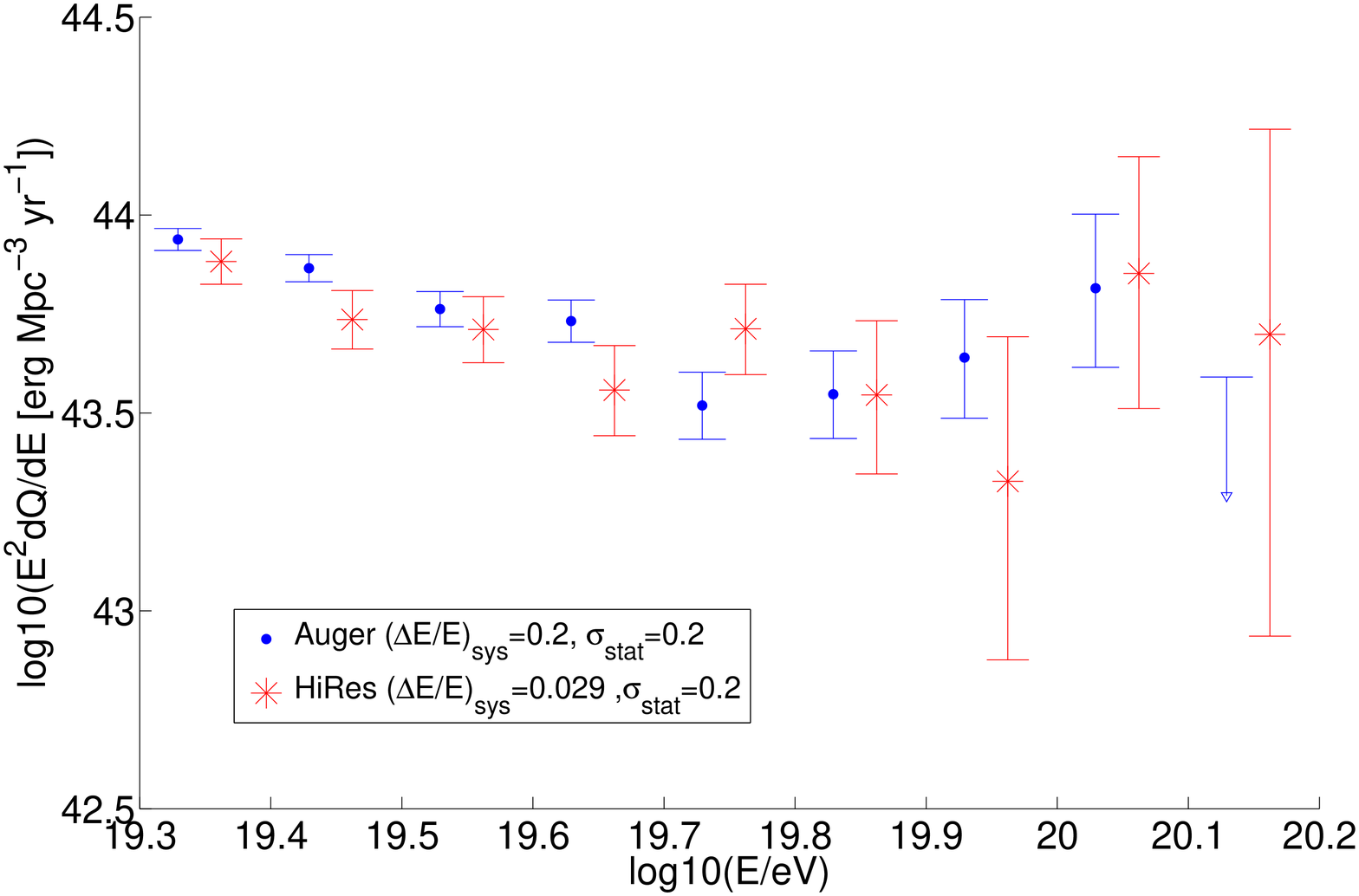} \caption{The energy generation rate as measured by Auger and Hires
assuming that the CRs are purely protons. Here the absolute energy scales of the Auger and Hires data where altered as
shown (in order that they agree at $\vep\sim 10^{19}\eV$, see \sref{sec:ReasonableFits}). In this figure we took into
account the statistic errors in the particle energy estimates according to eq.
\eqref{eq:smear}.}\label{fig:EnergyGenerationSmSh}
\end{figure}

To illustrate the effect of these uncertainties, we show in figure \ref{fig:EnergyGenerationSmSh} the energy generation rate implied by the data of Auger and Hires with
the Auger energy scale shifted by $20\%$ (within the reported systematic error, and as required in order that the two
experiments agree), and correcting for the statistic smearing using Eq.\eqref{eq:smear} with $\sig_{\text{stat}}=0.2$
for both experiments. By a fractional shift, $(\Delta
\vep/\vep)_{\text{sys}}$, in the absolute energy scale of an experiment, we mean that particles with a measured energy $\vep$ have
in fact an energy of $\vep+(\Delta \vep/\vep)_{\text{sys}}\vep$.

Another source of uncertainty is the unknown value of the spectral index $\alpha$. The spectral index can in
principle be directly read off from figures \ref{fig:EnergyGeneration} and \ref{fig:EnergyGenerationSmSh}. In practice, this is problematic due to the unknown Galactic and heavier nuclei contribution at energies $\vep\lesssim10^{19.5}\eV$ and due to the possible effect of the unknown local CR source density on the flux at particle energies $\vep\gtrsim 10^{20}\eV$, as explained above. This leaves a limited range of energies where the observed spectrum can be safely used to infer the generation spectrum, $10^{19.5}\eV\lesssim\vep\lesssim 10^{20}\eV$, where the limited statistics and the unknown dependence of $(\Delta \vep/\vep)_{\text{sys}}$ and $(\Delta \vep/\vep)_{\text{stat}}$ on $\vep$ does not allow an accurate determination of $\alpha$. The spectral index cannot be much smaller than $\alpha\approx 2$, as it would predict too many events at the highest energy and require a significant Galactic contribution at energies $\vep>10^{19.5}$. The spectral index is conservatively limited from above by $\alpha<2.7$, since larger values will produce a flux at lower energies, $\vep\sim10^{19}$~eV, that exceeds the observed flux (see more detailed discussion in \S~\ref{sec:RuleOut}).

A simple analytic estimate of the XG energy generation rate can be made as follows. In the energy range
$10^{19.2}\eV<\vep<10^{19.6}\eV$, slightly below the threshold for pion production by interaction with CMB photons, the
energy loss of protons is dominated by pair production and the effective generation time is approximately given by
$t_{eff}\approx(\tau_{0,ep}^{-1}+2H_0)^{-1}\approx 2.5\times10^{9}\yr$. The energy density of CRs per logarithmic
particle energy at $\vep = 10^{19.6}\eV$ is roughly $\vep^2dn/d\vep\approx 10^{-20.6}\erg\cm^{-3}$. Thus we can
approximate the energy production at $\vep\lesssim 10^{19.6}\eV$ as:
\begin{equation*}\label{eq:EnergyProduction}
\vep^2Q(\vep\lesssim10^{19.6}\eV)\approx (\alpha-1)\vep^2dn/d\vep|_{10^{19.6}\textrm{eV}} \times(\tau_{0,ep}^{-1}+2H_0)
\end{equation*}
\begin{equation}
\approx 0.3\times10^{44}(\alpha-1)\left(\frac{\vep^2dn/d\vep|_{10^{19.6}\textrm{eV}}}{10^{-20.6}\erg\cm^{-3}}\right)\erg\Mpc^{-3}\yr^{-1}.
\end{equation}

Note that there have been claims in the literature that the production rate is considerably higher or that the
uncertainty is much larger \citep[e.g.][]{Farrar08}. The source of the confusion is that different authors estimate the
production rate of XG CRs at different CR energies, e.g. at $\vep>10^{19}\eV$ (as in this ms) vs. $\vep\sim 10^{18}$
\citep[e.g.][]{Berezinsky08}. In particular, the generation rate per logarithmic particle energy in the model presented
by \citet{Berezinsky08} at energies $\vep\sim 10^{19.6}\eV$ and $\vep\sim 10^{18}\eV$ is
$\vep^2Q(\vep)|_{10^{19.6}\eV}\approx 1.3\times 10^{44}\erg\Mpc^{-3}\yr$ and $\vep^2Q(\vep)|_{10^{18}\eV}\approx 17\times
10^{44}\erg\Mpc^{-3}\yr$ respectively. The energy generation rate at $\vep> 10^{19}\eV$ is in agreement with the
generation rate estimates presented here [given that in Berezinsky's model $\alpha=2.7$ and $(\Delta
E/E)_{\text{sys}}\approx 40\%$ is required for the Auger energy calibration, in order that the Auger flux agrees with the
model at $\vep\sim 10^{19}\eV$, see \sref{sec:RuleOut}].

\section{Transition From Galactic to Extra Galactic CRs}\label{sec:Transition}

In this section we examine the CR spectrum down to lower energies and discuss different models that may explain the
smoothness of the spectrum from the knee to the highest observed energies. First we review in \sref{sec:FineTuning} the
argument that a transition between two different CR sources should be observed as a flattening in the spectrum, which
implies that the transition between Galactic and XG sources is likely to occur at the 'ankle'. We demonstrate that a model in
which the transition is at $\vep\ll10^{19}\eV$ involves fine tuning. In \sref{sec:ReasonableFits} we present two examples
of specific models which are consistent with the data (figure \ref{fig:ReasonableFit} and figure
\ref{fig:ReasonableFitCutoff}). Finally we show in \sref{sec:RuleOut} that the model presented in \citep{Berezinsky08} is
somewhat in disagreement with the latest published data from the Auger Experiment.

\subsection{The fine-tuning argument for a Galactic-XG transition at the 'ankle'}\label{sec:FineTuning}
It is established that CRs at the lowest energies, $\vep\sim 1\GeV$, are of Galactic origin and it is likely that the highest
energy, $\vep\gtrsim 10^{19.5}$, particles are XG. A basic feature that is expected at any transition
between two different sources of CRs is a flattening of the spectrum. This is expected since in order that the flux from
the source of the higher energy particles exceed the flux from the source of particles at lower energies, it's spectrum
must be flatter. It is quite remarkable that there is only one observed flattening in the CR spectrum throughout the span
of 10 orders of magnitudes of particle energies. This flattening, the so called 'ankle', is observed at an energy of
roughly $\vep\sim 10^{18.7}\eV$. In any model where the transition is at lower energies, it must occur without an
observed flattening. This requires fine tuning in both the spectra and amplitudes of the two contributions \citep[e.g. ][]{Hillas84}.

\begin{figure}[h]
\epsscale{1.6} \plotone{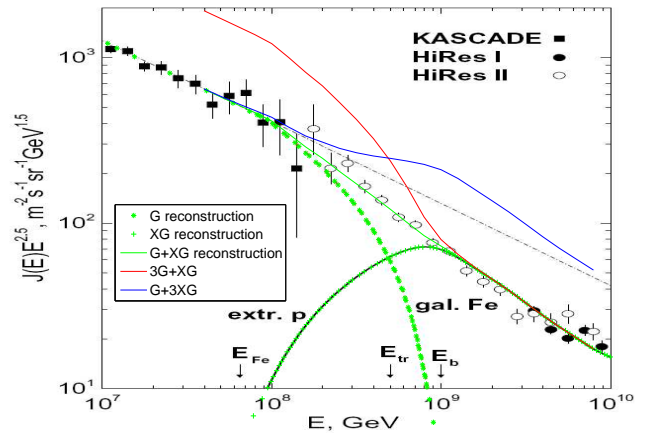} \caption{Demonstration of the fine tuning problem in models where the
transition from Galactic to XG CRs occurs without a flattening. The model presented in Figure 7 (left) from \citep{Berezinsky08}
is reproduced. The green asterisks, pluses and filled line are reproductions of the Galactic, XG and total fluxes respectively.
The red (blue) filled line is a slight modification of the model where the Galactic (XG) contribution was enhanced by a factor
of 3. As can be seen in both cases, a clear flattening that is not observed is expected in both
cases.}\label{fig:Ber08FineTuning}
\end{figure}

A class of such models pursued by some authors \citep[][and references therein]{Berezinsky08} suggests that the XG CRs
are dominant down to much lower energies, $\vep\sim 10^{18}\eV$. According to these models the flattening (ankle) is due
to the flattening ('dip') in the CR energy loss time as a function of energy. In these models the transition from Galactic to XG
sources has to occur without an observed flattening, which requires fine tuning between the parameters of the XG and Galactic components. To demonstrate this fine tuning we consider the model presented by \citet[][figure 7 left]{Berezinsky08}. A
scanned version of this figure is reproduced in figure \ref{fig:Ber08FineTuning}. On it, we added two slightly modified
models (the blue and red thin full lines) in which the Galactic and XG components were enhanced by a factor of 3 respectively.
As can be seen, there is a clear flattening that is expected in both cases at the transition region. Such flattening is
not observed.

It is important to stress that the amplitudes and spectra of the two contributions in this model are affected by many
physically unrelated factors. In particular, these have to be different types of accelerators (the Galactic ones limited in
energy and the XG capable of accelerating UHECRs). It is challenging to offer a physical reason for the spectra to adjust
in such a way as to erase the evidence for the transition.

\subsection{Two examples of models with a transition at the 'ankle'}\label{sec:ReasonableFits}

\begin{figure}[h]
\epsscale{1.4} \plotone{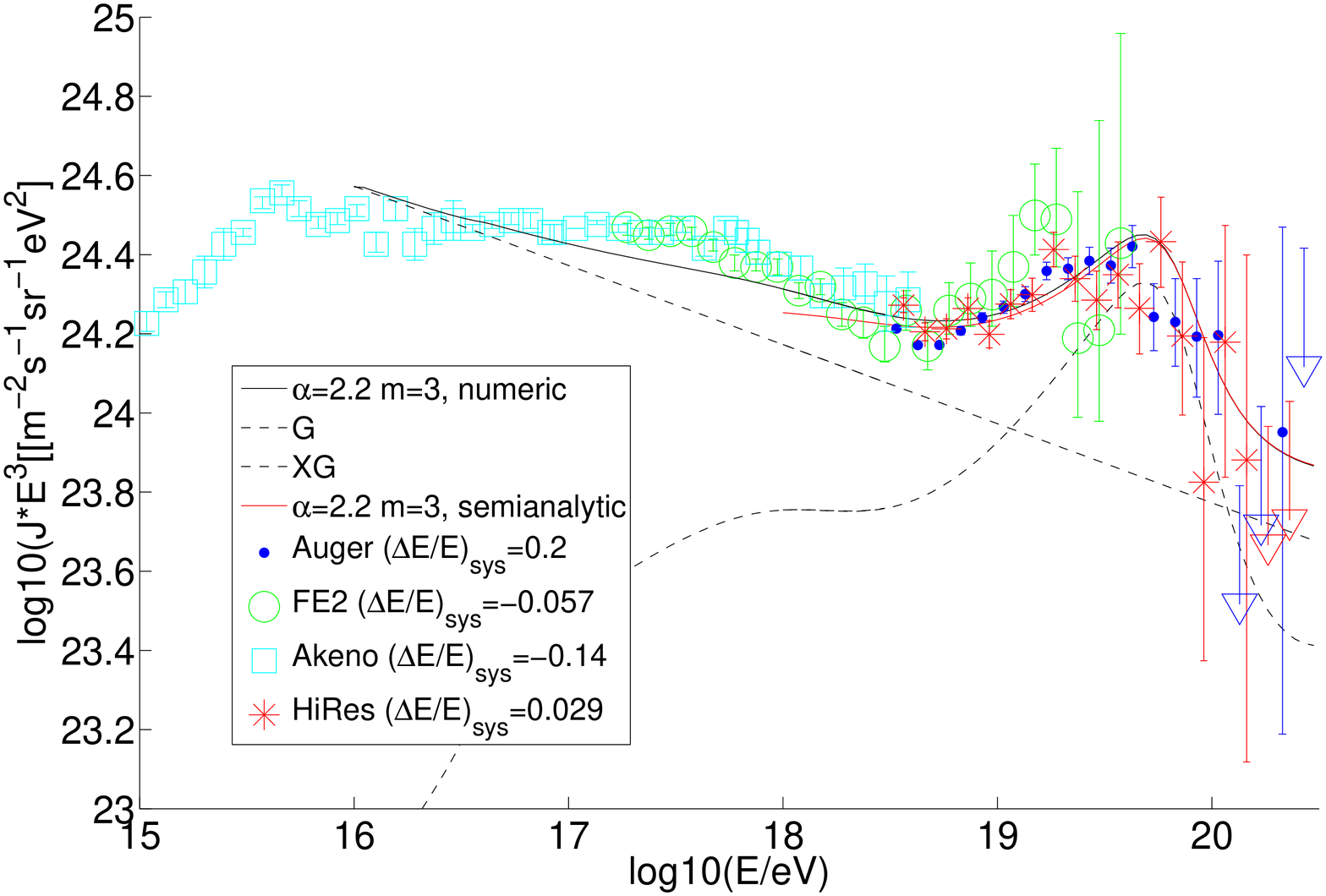} \caption{A model of UHECRs with a power law Galactic contribution [Eqs.
\eqref{eq:ReasonableFitG} and \eqref{eq:ReasonableFitXG}]. The expected flux is calculated numerically (full black line).
The expected flux using Eq. \eqref{eq:t_th} is shown for comparison for energies $\vep>10^{18}\eV$ (full red line). The
measured flux above the 'knee' as measured by Akeno \citep{Nagano00}, Fly's Eye \citep[][data taken from Nagano \& Watson
2000]{Bird94} and HiRes \citep{Abbasi08} is presented. The absolute energy calibration of the various experiments shown
were adjusted in order that they agree at $\vep\sim 10^{19}\eV$. The fractional shifts in the absolute energy scale,
$\Delta E/E$, are within the published systematic errors.}\label{fig:ReasonableFit}
\end{figure}

\begin{figure}[h]
\epsscale{1.4} \plotone{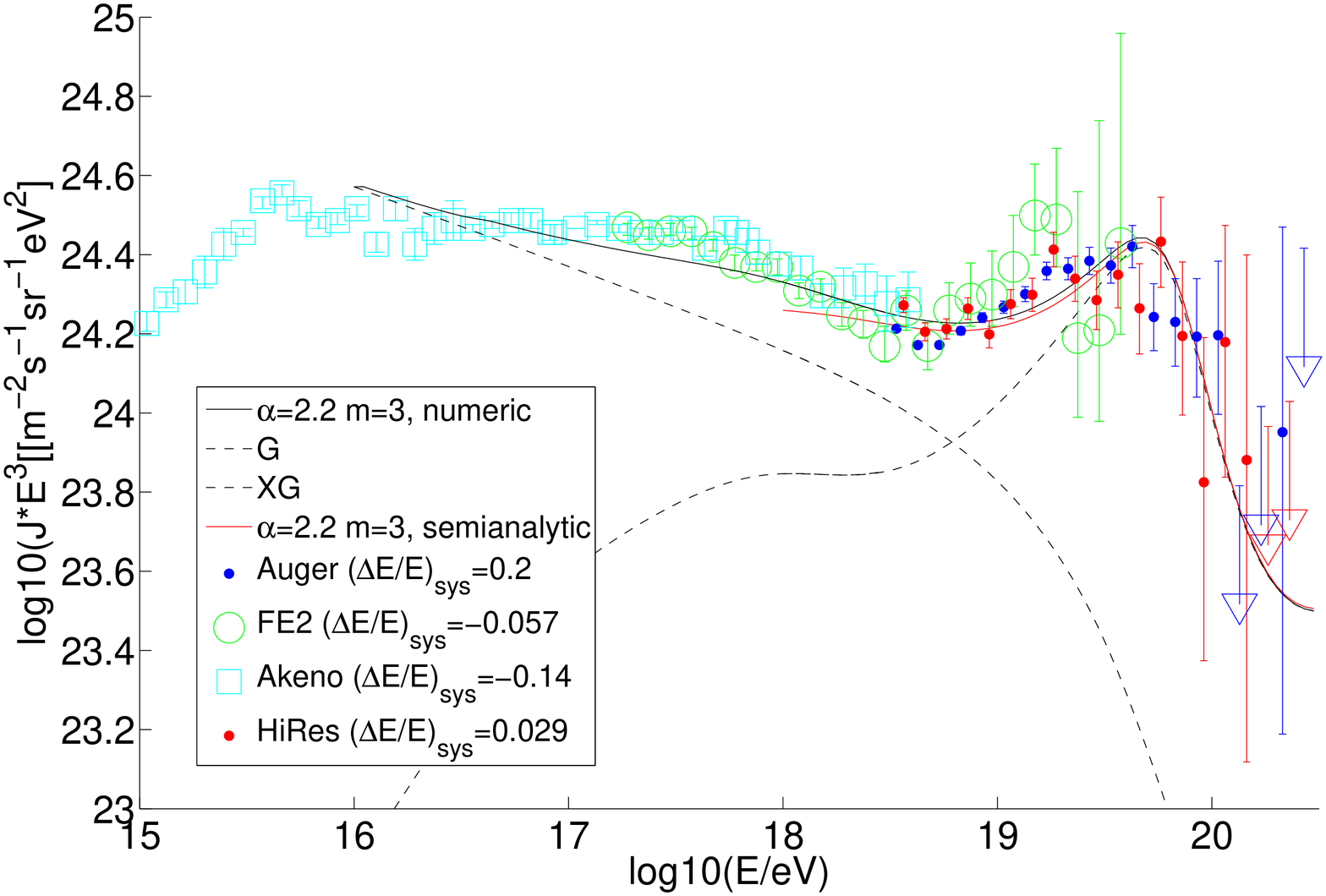} \caption{Same as figure \ref{fig:ReasonableFitCutoff}, with the Galactic spectrum a power law with an exponential cutoff, Eqs. \eqref{eq:ReasonableFitG} and
\eqref{eq:ReasonableFitXG}.}\label{fig:ReasonableFitCutoff}
\end{figure}

In figures \ref{fig:ReasonableFit} and \ref{fig:ReasonableFitCutoff} we present two examples of phenomenological models that include a Galactic and XG contribution that reproduce the observed spectra to a satisfying level.
In figure \ref{fig:ReasonableFit} the Galactic contribution is a pure power law. The parameters are:
\begin{equation}\label{eq:ReasonableFitG}
J_G=1.7\times 10^{-32}(\vep/10^{18.6}\eV)^{-3.2}\m^{-2}\sec^{-1}\sr^{-1}~\eV^{-1},
\end{equation}
\begin{equation}\label{eq:ReasonableFitXG}
\vep^{2}Q_{XG}(\vep)=0.45\times 10^{-44}(\vep/10^{19.6}\eV)^{-0.2}\erg\Mpc^{-3}\yr^{-1}.
\end{equation}
In figure \ref{fig:ReasonableFitCutoff} the Galactic contribution is a power law with an exponential cutoff at
$\vep_{cf}=10^{19.5}\eV$. The parameters are:
\begin{equation}\label{eq:ReasonableFitCuttoffG}
J_G=1.7\times 10^{-32}(\vep/10^{18.6}\eV)^{-3.2}\exp(-\vep/\vep_{cf})\m^{-2}\sec^{-1}\sr^{-1}~\eV^{-1},
\end{equation}
\begin{equation}\label{eq:ReasonableFitCuttoffXG}
\vep^{2}Q_{XG}(\vep)=0.55\times 10^{-44}(\vep/10^{19.6}\eV)^{-0.2}\erg\Mpc^{-3}\yr^{-1}.
\end{equation}

We make the following observations regarding these fits. These models agree with the data for energies $\vep>10^{16}\eV$
to about $20\%$. There are many factors that affect the fluxes and are not taken into account: Contribution from heavier
elements, diffusion in the intergalactic magnetic fields, non trivial contribution from nearby sources etc.  We believe
that without a better understanding of these systematics, using simple phenomenological models is the best one can do. On
the one hand, it is encouraging that simple models can relatively well explain the data on such a wide energy range. On
the other hand, it is clear that from studying the CR spectra it is very difficult to distinguish between the Galactic and XG
contributions. Thus, it is difficult to use these measurements to learn about the different unknown factors. The fact
that two basically different models agree with similar success is a good illustration of this problem.

\subsection{Can we already rule out some models based on the measured all particle spectrum?}\label{sec:RuleOut}
In order to use the all particle spectrum to rule out models, these models must have clear predictions for the spectrum.
As the chemical composition of the CRs is unknown and since there can be a Galactic contribution to high energies (see
\sref{sec:Transition}) the all particle spectrum by itself gives us limited information on the detailed XG UHECR
generation.

The so called 'dip' models \citep[][and references therein]{Berezinsky08}, which assume a pure proton composition with no
Galactic contribution at high energies, have a rather clear prediction for what the all particle spectrum should be. What makes
these models interesting, is the claimed agreement between the observed spectrum and the predicted spectrum based on this
assumption. Here we compare the model given in \citep{Berezinsky08} with the latest published data of the Auger
experiment. This model has a CR energy production rate of $\vep^2Q(\vep)= 16.7\times
10^{44}(\vep/10^{18}\eV)^{-0.7}\erg\Mpc^{-3}\yr^{-1}$ (this corresponds to a total CR emissivity of $L_0=3.7\times
10^{46}\erg\Mpc^{-3}\yr^{-1}$ for particle energies $\vep>1\GeV$, for a spectrum $Q\propto \vep^{-2}$ at
$1\GeV<\vep<10^{18}\eV$ and $Q\propto \vep^{-2.7}$ at $\vep>10^{18}\eV$). The spectral index and amplitude of this model
were obtained by fitting the data from several experiments in the energy range $\vep>10^{18}\eV$. As claimed by the
authors, the absolute energy calibration of the different experiments is fixed by the position of the 'dip' in the
spectrum and thus the spectral index and the amplitude can not change by much.

\begin{figure}[h]
\epsscale{1.4} \plotone{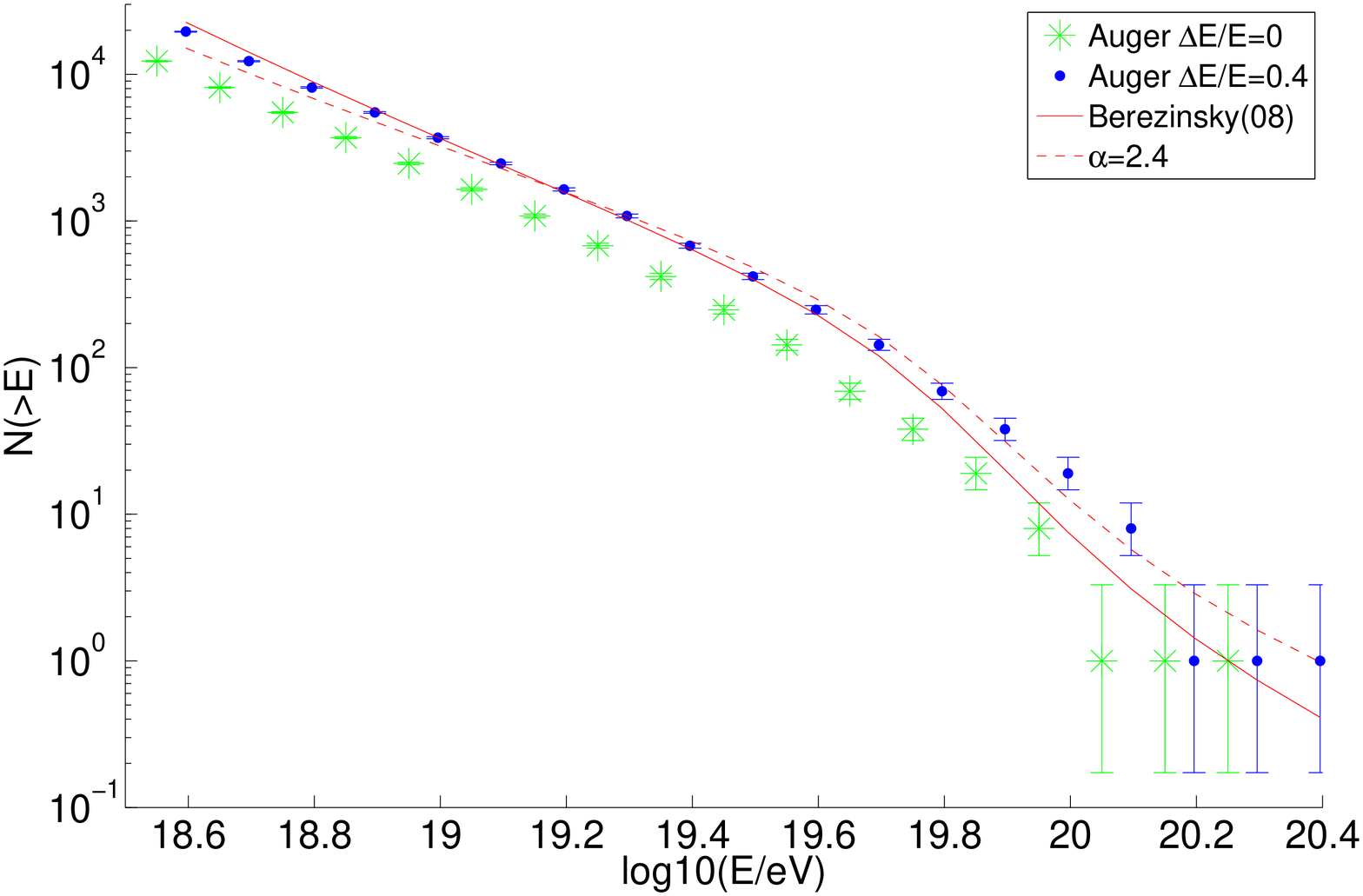} \caption{Cumulative number of events above different energies measured by the Auger experiment (green asterisks) compared to the expected number according to \citep{Berezinsky08} (red line). The Auger spectrum with the particles shifted by $40\%$ is shown on the figure (blue dots).}\label{fig:Ber08RuleOutStat}
\end{figure}

In figure \ref{fig:Ber08RuleOutStat} the cumulative number of events above different energies measured by the Auger
experiment (green asterisks) is compared to the expected number according to this model (red line). As can be seen, the
model over predicts the number of events \citep{Stanev08}. To achieve agreement at energies $\vep\sim10^{19}\eV$ it is necessary to assume
that the Auger experiment systematically underestimates the particle energies by about $40\%$. The spectrum that would
result if the energies of the particles were shifted by $40\%$ is shown on the figure (blue dots). It seems that the
shifted spectrum at the highest energies is flatter than expected by the model. For comparison, a model with a generation
spectrum of $Q\propto \vep^{-2.4}$ is plotted (red dashed line). It can be seen that even for this harder generation
spectrum, which is too hard to be consistent with the observed spectrum at $\vep\sim 10^{19}\eV$, the resulting flux
seems too soft to be consistent with the observed spectrum above $\vep\sim 10^{19.4}\eV$. We stress that in order that
this disagreement may be confidently used to rule this model out, the trend must be confirmed with higher statistics and
the systematics of the experiment must be better understood. In addition, an enhancement of the flux at the highest
energies, $\vep>10^{20}\eV$, may be a consequence of an enhancement in the local density of CR sources
\citep[e.g.][]{Giler80,Bahcall00,Waxman95b,Berezinsky08}.

\section{Summary and conclusions}\label{sec:Discussion}
In this paper we derived simple analytic tools for the analysis of the UHECR flux and used them to estimate the UHECR
generation rate.

We first derived in section \sref{sec:Model} approximate analytic expressions for the effective generation time
$t_{\text{eff}}(\vep)\equiv dn/d\vep /Q(\vep,z=0)$ of CRs, assuming the flux is dominated by protons coming from XG
sources. We used the fact that at high energies the energy loss time $\tau$ is much shorter than the Hubble time
$H_0^{-1}$ to derive approximate expressions for the effective generation time, Eqs.
\eqref{eq:general_first_order},~\eqref{eq:t_th_m_alpha_4_short}, and \eqref{eq:t_th}, by
expanding it to first order in the small parameter $H_0\tau$. We then obtained a phenomenological analytic approximation to
the loss time of protons due to pair production and pion production, Eq. \eqref{eq:proton_loss_approx},
\citep{Waxman95b}, and used it to derive a simple analytic expression for the effective generation time, Eq.
\eqref{eq:proton_effective_time}. We showed that these expressions are in good agreement with detailed numerical
calculations (figs. \ref{fig:t_th_m_alpha_4}, \ref{fig:t_th}, \ref{fig:t_th_m_alpha_4_analytic_Mpc} and
\ref{fig:t_th_m_alpha_0_3_2_3_analytic_Mpc}). In particular, for $2<\alpha<3$ and $0<m<3$ [see Eq. \eqref{eq:Q}], the
simple expression given in Eq. \eqref{eq:proton_effective_time} agrees with numerical calculations to better than $20\%$
at $\vep>10^{19}\eV$ as shown in figure \ref{fig:t_th_m_alpha_0_3_2_3_analytic_Mpc}.

Next, we estimated in \sref{sec:Production} the energy production rate of cosmic rays with energies $\varepsilon\gtrsim
10^{19.5}\eV$ using the latest reported measurements of the Auger experiment [figs \ref{fig:EnergyGeneration} and
\ref{fig:EnergyGenerationSmSh} and Eq. \eqref{eq:EnergyProduction}]. We showed that it is roughly energy and model
independent and equal to $\vep^2Q(\vep)|_{\vep\gtrsim 10^{19.5}\text{eV}}\sim
0.3-0.6\times10^{44}(\alpha-1)\erg\Mpc^{-3}\yr^{-1}$. This is consistent with earlier results derived by
\citet{Waxman95b} and \citet{Bahcall03}, and also by \citet{Berezinsky08}, who obtains
$\vep^2Q(\vep)\approx0.75\times10^{44}(\alpha-1)(\vep/10^{19.6}\eV)^{-\alpha+2}\erg\Mpc^{-3}\yr^{-1}$ for $\alpha=2.7$.
The slightly higher normalization obtained in the latter analysis is due mainly to the assumption, required by that analysis, that $(\Delta \vep/\vep)_{\text{sys}}\simeq40\%$ for Auger (see \sref{sec:RuleOut}), a systematic error which is larger than that quoted by the experiment \citep[note, that the considerably
higher CR energy production rate sometimes quoted from][refers to lower particle energies]{Berezinsky08}.

The statistic and systematic errors in the generation rate estimate are of the order of tens of percents. The main source for the systematic error is a $\sim 20\%$ systematic uncertainty in the energy calibration of the CR particle energies, which implies a $\sim 30\%$ uncertainty in the measured flux.
Another source of uncertainty is the unknown value of the spectral index $\alpha$. There is a limited range of energies where the observed spectrum can be safely used to infer the generation spectrum, $10^{19.5}\eV\lesssim\vep\lesssim 10^{20}\eV$, where the limited statistics and the unknown dependence of $(\Delta \vep/\vep)_{\text{sys}}$ and $(\Delta \vep/\vep)_{\text{stat}}$ on $\vep$ does not allow an accurate determination of $\alpha$. $\alpha$ is roughly limited to the range $2\lesssim\alpha<2.7$ (see discussion in \sref{eq:EnergyProduction} and \S~\ref{sec:RuleOut}).

In section \sref{sec:Transition} we showed that the measured flux of CRs at energies $\vep>10^{16}\eV$ is consistent with simple models that include a simple Galactic contribution and a transition between the Galactic and XG dominated regions at the 'ankle', $\vep\sim 10^{19}\eV$ (figs \ref{fig:ReasonableFit} and \ref{fig:ReasonableFitCutoff}). As explained in section \sref{sec:FineTuning} and illustrated in figure \ref{fig:Ber08FineTuning}, models in which the transition between Galactic and XG sources occurs at energies considerably below the ankle require a fine tuning in the amplitudes of the Galactic and XG contributions in order that the expected flattening in the transition will not be observed. Moreover, such models require a steep spectrum, $\alpha\simeq2.7$, which is disfavored by the data (see fig.~\ref{fig:Ber08RuleOutStat}). In fact, even harder generation spectrum, $\alpha=2.4$ , which is too hard to be consistent with the observed spectrum at $\vep\sim 10^{19}\eV$,
seems too soft to be consistent with the observed spectrum above $\vep\sim 10^{19.4}\eV$ (fig.~\ref{fig:Ber08RuleOutStat}). Nevertheless, it should be noted that in order that
this disagreement may be confidently used to rule out models with Galactic to XG transition well below the ankle, the trend must be confirmed with higher statistics and
the systematics of the experiment must be better understood. In addition, an enhancement of the flux at the highest
energies, $\vep>10^{20}\eV$, may be a consequence of an enhancement in the local density of CR sources
\citep[e.g.][]{Giler80,Bahcall00,Waxman95b,Berezinsky08}.

To conclude, we summarize what we believe the main conclusions that can be drawn from the analysis of the all particle
spectrum are.
\begin{enumerate}
\item The detection of the GZK cutoff \citep{Bahcall03,Abbasi08,Abraham08} supports the claim that UHECRs with energies $\vep\gg10^{19}\eV$ are mainly protons arriving from XG sources.
We note that the fact that we obtained a reasonable, smooth, CR generation spectrum (figs \ref{fig:EnergyGeneration}
and \ref{fig:EnergyGenerationSmSh}) after correcting for the particle energy losses (which are strongly energy dependent)
is consistent with the presence of the GZK cutoff.
\item Assuming UHECRs are XG protons, the energy generation rate is $\vep^2Q(\vep\gtrsim10^{19.5}\eV)\sim 0.3-0.6\times10^{44}(\alpha-1)\erg\Mpc^{-1}\yr^{-1}$ [see figs \ref{fig:EnergyGeneration} and \ref{fig:EnergyGenerationSmSh} and Eq. \eqref{eq:EnergyProduction}].
\item The fact that the only flattening observed in the spectrum is at the 'ankle' strongly suggests that the transition from Galactic to XG CRs occurs at that energy scale (see discussion in \sref{sec:FineTuning}), $\sim10^{19}\eV$.
\item The data in the energy range $\vep>10^{16}\eV$ is consistent with simple models including Galactic and XG contributions (the latter with a flat energy generation spectral index, $\alpha\simeq2$) where the transition occurs at the 'ankle' (figs \ref{fig:ReasonableFit},\ref{fig:ReasonableFitCutoff}).
\item The all particle spectrum alone cannot be used to discern the detailed spectral shapes of the Galactic and XG contributions (see \sref{sec:ReasonableFits}).  There is a limited range of energies where the observed spectrum can be safely used to infer the XG generation spectrum, $10^{19.5}\eV\lesssim\vep\lesssim 10^{20}\eV$, where the limited statistics and the unknown dependence of $(\Delta \vep/\vep)_{\text{sys}}$ and $(\Delta \vep/\vep)_{\text{stat}}$ on $\vep$ allow a broad range for the spectral index, $2\lesssim\alpha<2.7$.
\end{enumerate}

\acknowledgements
This research was partially supported by ISF, AEC, and Minerva grants.

\appendix

\section{The effect of statistic errors in the measurements of individual particle energies on the derived flux.}\label{sec:Smear}
Suppose that each particle with energy $\vep_{0}$ has a probability
$P(\frac{\vep-\vep_{0}}{\vep_{0}})\frac{d\vep}{\vep_{0}}$ to be measured to carry an energy $\vep$ in the interval
$\vep,\vep+d\vep$.
%$x=\frac{\vep-\vep_{0}}{\vep_{0}}$ so $\vep_{0}=\frac{\vep}{1+x}$
The number of particles measured in the interval $[\vep_{1},\vep_{2}]$ is
\begin{equation}\label{eq:statistics1}
N(\vep_{1}<\vep<\vep_{2})=\int_{\vep_{1}}^{\vep_{2}}\int_{0}^{\infty}d\vep_{0}\frac{dN}{d\vep}(\vep_0)P(\frac{\vep-\vep_{0}}{\vep_{0}})\frac{d\vep}{\vep_{0}},
\end{equation}
%\begin{equation}
%\int_{\vep_{1}}^{\vep_{2}}\frac{d\vep}{\vep}\int_{-1}^{\infty}\frac{d\vep_{0}\vep}{\vep_{0}^{2}}\frac{dN}{d\vep_{0}}|_{(\frac{\vep}{1+x})}\vep_{0}P(x)dx=
%\end{equation}
where $dN/d\vep$ is the number of particles per unit energy that reached the detector during the measurement. By changing
variables $[\vep,\vep_0]$ to $[\vep, x=\frac{\vep-\vep_{0}}{\vep_{0}}]$ we can rewrite this expression as (after some
straight forward algebra)
\begin{equation}
\int_{\vep_{1}}^{\vep_{2}}d\vep\int_{-1}^{\infty}dx\frac{dN}{d\vep}(\frac{\vep}{1+x})\frac{1}{1+x}P(x)dx.
\end{equation}
Assuming that the distribution of arriving particles can be described locally (in energy) as a power law distribution
$dN/d\vep=A\vep^{-\beta}$, where the value of $\beta$ can be derived from the derivatives of the distribution in the
energy regime under consideration, $\beta=d\log(N)/d\log(\vep)$ we find [using Eq. \eqref{eq:statistics1}]
\begin{equation}
N(\vep_{1}<\vep<\vep_{2})=A\int_{\vep_{1}}^{\vep_{2}}d\vep\vep^{-\beta}\int_{-1}^{\infty}dx(1+x)^{\beta-1}P(x)dx=N_{0}(\vep_{1}<\vep<\vep_{2})\int_{-1}^{\infty}dx(1+x)^{\beta-1}P(x)dx,
\end{equation}
where $N_{0}(\vep_{1}<\vep<\vep_{2})$ is the true number of particles that reached the detectors in the range
$\vep_1<\vep<\vep_2$. Assuming $\sig_{\text{stat}}\equiv \ave{x^2}\ll1$ and that the energies are calibrated so that
$\ave{\vep}=\vep_0$, we get
\begin{equation}\label{eq:statistics_approx}
N(\vep_{1}<\vep<\vep_{2})=N_{0}(\vep_{1}<\vep<\vep_{2})(1+\frac{(\beta-1)(\beta-2)}{2}\sig_{\text{stat}}^2+O(\sig_{\text{stat}}^4)).
\end{equation}

The value of $\beta$ can be written in terms of the production and propagation as $\beta=\alpha+\beta_{prop},$ where
$\alpha=-d\log(Q)/d\log\vep$ and
\begin{equation}
\beta_{prop}=-d\log(t_{eff})/d\log\vep.
\end{equation}
For propagation of protons, $\beta_{prop}$ reaches approximately $3$ at $\vep\approx 10^{20}$. For $2<\alpha<3$, $\beta$
can be as large as $5$ to $6$. For a value of $\sig_{\text{stat}}\approx 0.2$ typical to these experiments
\citep[e.g.][]{Abraham08} this implies an error of $30-40\%$ at peak ($\vep\sim 10^{20}\eV$). If higher precision in the
flux estimate is required, an estimate of $\sigma$ at these energies is essential.

At lower energies $\vep<10^{19.5}$, we measure $\beta\approx 3$ directly and the errors are of the order of a few
percents.

\end{document}

%% file: Definitions.tex
\newcommand{\DDir}{\relax{D\kern-.7em{/}}}

\newcommand{\inv}[1]{\frac{1}{#1}}

\newcommand{\be}{\begin{equation}}
\newcommand{\ee}{\end{equation}}
\newcommand{\bea}{\begin{equation*}}
\newcommand{\eea}{\end{equation*}}

\newcommand{\ave}[1]{\left\langle #1\right\rangle}

\newcommand{\pr}{\partial}

\newcommand{\nin}{\relax{\in\kern-.8em{/}}}

\newcommand{\sig}{\sigma}

\newcommand{\vep}{\varepsilon}

\newcommand{\cm}{\mbox{ cm}}
\newcommand{\m}{\mbox{ m}}
\newcommand{\sr}{\mbox{ sr}}

\newcommand{\yr}{\mbox{ yr}}

\newcommand{\erg}{\mbox{ erg}}

\newcommand{\km}{\mbox{ km}}

\newcommand{\kpc}{\mbox{ kpc}}
\newcommand{\Mpc}{\mbox{ Mpc}}
\newcommand{\eV}{\mbox{eV}}

\newcommand{\GeV}{\mbox{~GeV}}

\newcommand{\sref}{\S~\ref}